\documentclass[a4paper,UKenglish]{lipics-v2016}

\usepackage{microtype}%if unwanted, comment out or use option "draft"

\usepackage{tikz}
\usepackage{amsmath}
\usepackage{tabu}
\usepackage{multirow}
\usepackage{booktabs}
\usepackage{color}
\usepackage[binary-units]{siunitx}

\usepackage{graphicx}
\graphicspath{ {./img/} }
\usepackage{subcaption}

\DeclareMathOperator{\CSA}{CSA}
\DeclareMathOperator{\Oh}{\mathcal{O}}

\DeclareMathOperator{\ftSearch}{t_{search}(\vsPat)}
\DeclareMathOperator{\ftLookup}{t_{lookup}(\vsTColl)}

\newcommand\vColl{\mathcal{D}}
\newcommand\vDoc{\mathit{T}}
\newcommand\vnDocs{\mathit{d}}
\newcommand\vTColl{\mathcal{T}}
\newcommand\vsTColl{\mathit{n}}
\newcommand\vTCollSep{\mathit{\$}}
\newcommand\vPat{\mathit{P}}
\newcommand\vsPat{\mathit{m}}

\newcommand\occ{\mathit{occ}}
\newcommand\ndoc{\mathit{ndoc}}

\newcommand\SA{\mathit{SA}}
\newcommand\DA{\mathit{DA}}
\newcommand\vSPosPat{\mathit{sp}}
\newcommand\vEPosPat{\mathit{ep}}

\newcommand\vILCP{\mathit{ILCP}}
\newcommand\vnRuns{\rho}
\newcommand\vGrm{\mathcal{G}}
\newcommand\vnrules{\mathit{g}}

\newcommand\vsBlock{\mathit{b}}
\newcommand\vStorFac{\beta}

\newcommand\idxName{\textsf}

\renewcommand{\log}{\lg}

\title{Fast, Small, and Simple Document Listing on Repetitive Text Collections%
\footnote{Funded with basal funds FB0001 and by Fondecyt Grant 1-170048, Conicyt, Chile.}}

\author{Dustin Cobas}
\author{Gonzalo Navarro}

\affil{CeBiB --- Center for Biotechnology and Bioengineering,
Department of Computer Science, University of Chile, Chile. 
{\tt dustin.cobas@gmail.com,gnavarro@dcc.uchile.cl}}

\Copyright{Dustin Cobas, Gonzalo Navarro}



\begin{document}

\maketitle

\begin{abstract}
Document listing on string collections is the task of finding all documents 
where a pattern appears. It is regarded as the most fundamental document
retrieval problem, and is useful in various applications. Many of the 
fastest-growing 
string collections are composed of very similar documents, such as versioned 
code and document collections, genome repositories, etc. Plain pattern-matching
indexes designed for repetitive text collections achieve orders-of-magnitude
reductions in space. Instead, there are not many analogous indexes for document
retrieval. In this paper we present a simple document listing index for 
repetitive string collections of total length $n$ that lists the $\ndoc$ 
distinct documents where a pattern of length $m$ appears in time 
$\Oh(m+\ndoc \cdot \log n)$. We exploit the repetitiveness of the {\em document 
array} (i.e., the suffix array coarsened to document identifiers) to 
grammar-compress it while precomputing the answers to nonterminals, and store 
them in grammar-compressed form as well. Our experimental results show that our 
index sharply outperforms existing alternatives in the space/time tradeoff map.
\end{abstract}

\section{Introduction}
\label{sec:intro}

{\em Document retrieval} is a family of problems aimed at retrieving {\em 
documents} from a set that are relevant to a query {\em pattern}. In a rather
general setting, both documents and patterns are arbitrary strings. This
encompasses the well-known application of natural language and Web searching,
but also many others of interest in bioinformatics, software development, and
multimedia retrieval, to name a few \cite{Nav14}.

The most fundamental document retrieval problem, on top of which more
sophisticated ones are built, is {\em document listing}. This problem aims
at simply returning the list of documents where the pattern appears.
An obvious solution to document listing resorts to {\em pattern matching}:
find all the $\occ$ positions where the pattern appears, and then return the
$\ndoc$ different documents where those positions lie. This solution requires time
$\Omega(\occ)$ and the output is of size $\Oh(\ndoc)$, so the approach is very 
inefficient if $\ndoc \ll \occ$ (i.e., if the pattern appears many times in the
same documents). A better solution, which however applies only in natural 
language settings, resorts to {\em inverted indexes} \cite{BYRN11}. These restrict 
the possible patterns to sequences of words and store the list of the documents 
where each word appears, thereby solving document listing via intersections of 
the lists of the pattern words.

Muthunkishnan \cite{Mut02} designed the first linear-space and optimal-time
index for general string collections. 
Given a collection of total length $n$, he builds an index of $\Oh(n)$
words that lists the $\ndoc$ documents where a pattern of length $m$ appears
in time $\Oh(m+\ndoc)$. While linear space is deemed as sufficiently small in
classic scenarios, the solution is impractical for very large text collections
unless one resorts to disk, which is orders of magnitude slower.
Sadakane \cite{Sad07} showed how to reduce the space of Muthukrishnan's index
to that of the statistically-compressed text plus $\Oh(n)$ {\em bits}, while
raising the time complexity to only $\Oh(m+\ndoc\cdot \log n)$ if the appropriate 
underlying pattern-matching index is used \cite{BN13}.

The sharp growth of text collections is a concern in many recent applications,
outperforming Moore's Law in some cases \cite{Plos15}. 
Fortunately, many of the fastest-growing text
collections are {\em highly repetitive}: each document can be obtained from 
a few large blocks of other documents. These collections arise in different
areas, such as repositories of genomes of the same species (which differ from
each other by a small percentage only) like the 100K-genome
project\footnote{{\tt https://www.genomicsengland.co.uk/about-genomics-england/the-100000-genomes-project}}, software repositories that store all the versions 
of the code arranged in a tree or acyclic graph like GitHub\footnote{{\tt 
https://github.com/search?q=is:public}}, versioned document repositories where 
each document has a timeline of versions like Wikipedia\footnote{{\tt 
https://en.wikipedia.org/wiki/Wikipedia:Size\_of\_Wikipedia}}, etc.
On such text collections, statistical compression is ineffective \cite{KN13} 
and even $\Oh(n)$ bits of extra space can be unaffordable. 

Repetitiveness is the key to tackle the fast growth of these collections: their
amount of new material grows much slower than their size. For example, version
control systems compress those collections by storing the list of edits with 
respect to some reference document that is stored in plain form, and reconstruct
it by applying the edits to the reference version. Much more challenging,
however, is to {\em index} those collections in small space so as to support 
fast pattern matching or document retrieval tasks. To date, there exist several
pattern matching indexes for repetitive text collections (see a couple of
studies \cite{Nav12,GNP18} and references therein). However, there are not 
many document retrieval indexes for repetitive text collections 
\cite{CM13,GHKKNPS17,Nav17}. Most of these indexes \cite{Sad07,GHKKNPS17} rely
on a pattern-matching index needs $\Omega(n)$ bits in order to offer
$\Oh(\log n)$ time per retrieved document.

In this paper we introduce new simple and efficient document listing indexes
aimed at highly repetitive text collections. Like various preceding indexes, 
we achieve $\Oh(m+\ndoc\cdot\log n)$ search time, yet our indexes are way
faster and/or smaller than previous ones on various repetitive datasets,
because they escape from the space/time tradeoff of the pattern-matching index.
Our main idea is as follows: we use the {\em document array} $\DA[1..n]$ 
\cite{Mut02}, which projects the entries of the {\em suffix array} \cite{MM93} 
to the document where each position belongs. Document listing boils down to 
listing the distinct integers in a range $\DA[sp..ep]$, where $sp$ and $ep$ 
are found in time $\Oh(m)$. Array $\DA$ must be grammar-compressible since the 
differential suffix array is known to be so on repetitive texts 
\cite{GNF14,GNP18}. We then build a {\em balanced} binary context-free grammar 
that generates (only) $\DA$. This allows us retrieve any individual cell of
$\DA$ in time $\Oh(\log n)$ and any range $\DA[sp..ep]$ in time 
$\Oh(ep-sp+\log n)$. We can then implement existing indexes 
\cite{Sad07,GHKKNPS17} within much less space and without affecting their
time complexities. Further, we propose a new simple index based on the
grammar-compressed array $\DA$. Our compression guarantees that any range 
$\DA[sp..ep]$ is covered by $\Oh(\log n)$ nonterminals. For each nonterminal 
of the grammar, we store the list of the distinct documents appearing in it. 
The set of all the lists is grammar-compressed as well, as done in previous 
work \cite{CM13,GHKKNPS17}. We then merge the lists of the $\Oh(\log n)$ 
nonterminals that cover $\DA[sp..ep]$, in time $\Oh(\ndoc\cdot\log n)$.

\section{Preliminaries}

%\subsection{Notation}

A {\em document} $\vDoc$ is a sequence of symbols over an alphabet $\Sigma = 
[1..\sigma]$, terminated by a special symbol $\vTCollSep$ that is 
lexicographically smaller than any symbol of $\Sigma$.

A {\em collection} $\vColl$ is a set of $\vnDocs$ documents $\vColl = \{\vDoc_1, \dots, \vDoc_{\vnDocs}\}$.
$\vColl$ is commonly represented as the concatenation of all its documents, 
$\vTColl = \vDoc_1 \vDoc_2 \dots \vDoc_\vnDocs$, of length $|\vTColl|=\vsTColl$.

A {\em pattern} $\vPat$ is a string over the same alphabet $\Sigma$ with length $|\vPat| = \vsPat$.
It occurs $\occ$ times in $\vTColl$, and appears in $\ndoc$ documents.

\vspace*{-4mm}
\subparagraph{Text indexes. \rm
The {\em suffix tree} \cite{Wei73} of a string $\vTColl$ is a compressed digital tree storing all the suffixes $\vTColl[i..\vsTColl]$, for all $1 \leq i \leq \vsTColl$. The suffix tree node reached by following the symbols of a pattern $P$
is called the {\em locus} of $P$ and is the ancestor of all the $\occ$ leaves 
corresponding to the positions of $P$ in $\vTColl$. The suffix tree uses
$\Oh(n\log n)$ bits and lists all the occurrences of $\vPat$ in 
time $\Oh(\vsPat + \occ)$. \\}

The {\em suffix array} \cite{MM93} $\SA[1..n]$ of a string $\vTColl[1..n]$ is a permutation of the starting positions of all the suffixes of $\vTColl$ in lexicographic order, $\vTColl[\SA[i], \vsTColl] < \vTColl[\SA[i + 1], \vsTColl]$ for all $1 \leq i < \vsTColl$.
$\SA$ can be binary searched to obtain the range $\SA[\vSPosPat.. \vEPosPat]$ of all the suffixes prefixed by $\vPat$ (note $\occ = ep-sp+1$). Thus the occurrences of $P$ can
be listed in time $\Oh(\vsPat \lg \vsTColl + \occ)$.
The suffix array takes $\vsTColl \lg \vsTColl$ bits.

\emph{Compressed suffix arrays} ($\CSA$s) \cite{NM06} are space-efficient representations of the suffix array.
They find the interval $[\vSPosPat.. \vEPosPat]$ corresponding to $\vPat[1..m]$ in time $\ftSearch$, and access any cell $\SA[i]$ in time $\ftLookup$.
Their size in bits, $|\CSA|$, is usually bounded by $\Oh(\vsTColl \lg \sigma)$.

\vspace*{-4mm}
\subparagraph{Grammar compression. \rm
Grammar compression of a string $S[1..n]$ replaces it by a context-free grammar (CFG) $\vGrm$ that uniquely generates $S$.
This CFG $\vGrm$ may require less space than the original sequence $S$,
especially when $S$ is repetitive. \\ }
Finding the smallest CFG $\vGrm^*$ generating the input $S$ is NP-hard 
\cite{LS02}, but various $\Oh(\lg (n / |\vGrm^*|))$-approximations
exist. In particular, we are interested in approximations that are 
{\em binary} (i.e., the maximum arity of the parse tree is 2) and
{\em balanced} (i.e., any substring is covered by $\Oh(\log n)$ maximal 
nodes of the parse tree) 
\cite{Ryt03,CLLPPSS05,Jez16}.

\section{Related Work}\label{rel_work}

%Muthu
Muthukrishnan \cite{Mut02} proposed the first optimal-time linear-space solution
to the document listing problem. He defines the \emph{document array} 
$\DA[1..n]$ of $\vTColl$, where $\DA[i]$ stores the identifier of the document 
to which $\vTColl[\SA[i]]$ belongs. The document listing problem is then
translated into computing the 
$\ndoc$ distinct identifiers in the interval $\DA[\vSPosPat.. \vEPosPat]$ 
corresponding to the pattern $P$. He uses a suffix tree to find $sp$ and $ep$
in time $\Oh(m)$, and then an algorithm that finds the $\ndoc$ distinct numbers
in the range in time $\Oh(\ndoc)$.

%Sada
Sadakane \cite{Sad07} adapts the method of Muthukrishnan to use much
less space. He replaces the suffix tree by a CSA, and mimics the algorithm to
find the distinct numbers in $\DA[sp..ep]$ using only $\Oh(n)$ bits of space.
Within $|\CSA|+\Oh(n)$ bits, he performs document listing in time 
$\Oh(\ftSearch + \ndoc \cdot \ftLookup)$. Using a particular CSA \cite{BN13}
the space is $n\log\sigma + o(n\log\sigma) + \Oh(n)$ bits and the time is
$\Oh(m + \ndoc \cdot \log n)$.

There are many other classical and compact indexes for document listing. We
refer the reader to a survey \cite{Nav14} and focus on those aimed at
repetitive text collections.

%ILCP
Gagie et al.~\cite{GHKKNPS17} proposed a technique adapting Sadakane's
solution to highly repetitive collections. They show that the technique to find
the distinct elements of $\DA[sp..ep]$ can be applied almost verbatim on an 
array they call \emph{interleaved longest-common-prefix array} ($\vILCP$).
On repetitive collections, this array can be decomposed into a small number
$\vnRuns$ of equal values, which allows them represent it in little space.
The {\em ILCP} index requires $|\CSA| + \Oh((\vnRuns + \vnDocs)\lg\vsTColl)$
bits of space and solves document listing in time $\Oh(\ftSearch + \ndoc \cdot \ftLookup)$.

%PDL
Gagie et al.~\cite{GHKKNPS17} proposed another radically different approach, 
called {\em Precomputed Document Lists (PDL)}. The idea is to store the 
list of the documents where (the corresponding substring of) each suffix tree 
node appears. Then the search consists of finding the
locus of $P$ and returning its list. To reduce space, however, only some sampled
nodes store their lists, and so document listing requires {\em merging} the
lists of the maximal sampled nodes descending from the locus node. To
further save space, the lists are grammar-compressed, which is 
effective when the collection $\vColl$ is repetitive.% \cite{NavarroPV14:GDR}.

To bound the query time, the deepest sampled nodes cover at most $\vsBlock$ 
leaves, and a factor $\vStorFac$ restricts the work done per document
to be merged in the unions of the lists.
The index then requires $|\CSA| + \Oh((\vsTColl / \vsBlock)\lg \vsTColl)$ 
bits, and the document listing time is $\Oh(\ftSearch + \ndoc \cdot \vStorFac \cdot h + \vsBlock \cdot \ftLookup)$, where $h$ is the height of the suffix tree.

%CSAs
A problem in all the revisited $\CSA$-based solutions are the extra 
$\Theta((\vsTColl \lg \vsTColl) / \ftLookup)$ bits that must be included
in $|\CSA|$ in order to get $\Theta(\ftLookup)$ time per document. This space does
not decrease with repetitiveness, forcing all these indexes to use $\Omega(n)$ 
bits to obtain time $\Oh(\ftSearch + \ndoc \cdot \log n)$, for example.

%Grammar
Claude and Munro \cite{CM13} propose the first index for document listing based
on grammar compression, which escapes from the problem above. They extend a
grammar-based pattern-matching index \cite{CN10} by storing the list 
of the documents where each nonterminal appears. Those lists are
grammar-compressed as well.
The index searches for the minimal nonterminals that contain $P$ and merges
their lists. While it does not offer relevant space or query time
guarantees, the index performs well in practice. Navarro \cite{Nav17} extends
this index in order to obtain space guarantees and 
$\Oh(m^2 + m\log^2 n)$ time, but the
scheme is difficult to implement.

\section{Our Document Listing Index}

Like most of the previous work, we solve the document listing problem by 
computing the $\ndoc$ distinct documents in the interval 
$\DA[\vSPosPat.. \vEPosPat]$ corresponding to the pattern $P$, found with a 
$\CSA$ in 
time $\Oh(\ftSearch)$. Instead of also using the $\CSA$ to compute the values
of $\DA$ (and thus facing the problem of using $\Theta((n\log n)/\ftLookup)$ 
bits to compute a cell in time $\Theta(\ftLookup)$, as it happens in previous work
\cite{Sad07,GHKKNPS17}), we store the array $\DA$ directly,
yet in grammar-compressed form. This is promising because the suffix array of
repetitive collections is known to have large areas $\SA[i..i+\ell]$ that 
appear shifted by 1 elsewhere, $\SA[j..j+\ell]$, that is, $\SA[i+k]=\SA[j+k]+1$
for all $0 \le k \le \ell$ \cite{MNSV09,GNP18}. 
Except for the $d$ entries of $\SA$ that point to
the ends of the documents, it also holds that $\DA[i+k] = \DA[j+k]$. Grammar
compression is then expected to exploit those large repeated areas in $\DA$.

To answer the queries efficiently, we use an idea similar to the one introduced
in PDL \cite{GHKKNPS17} and the Grammar-index \cite{CM13}: precomputing 
and storing the answers of document listing queries, and grammar-compressing
those lists as well. An important difference with them is that PDL stores
lists for suffix tree nodes and the Grammar-index stores 
lists for nonterminals of the grammar of $\vTColl$. Our index, 
instead, stores lists for the nonterminals of the grammar of
$\DA$. This is much simpler because we do not store a suffix tree
topology (like PDL) nor a complex grammar-based pattern-matching index (like the
Grammar-index): we simply find the interval $\DA[sp..ep]$ using the $\CSA$, fetch
the nonterminals covering it, and merge their lists. By 
using a binary balanced grammar on $\DA$, we ensure that any 
document is obtained in the merging only $\Oh(\log n)$ times, which leads
to our worst-case bound of $\Oh(\ndoc\cdot\log n)$. PDL and the Grammar-index
cannot offer such a logarithmic-time guarantee.

\subsection{Structure}

The first component of our index is a $\CSA$ suitable for repetitive 
collections, of which we are only interested in the functionality of finding
the interval $\SA[sp..ep]$ corresponding to a pattern $P[1..m]$. For example,
we can use the Run-Length CSA (RLCSA) variant of Gagie et al.~\cite{GNP18}, 
which offers 
times $\ftSearch = \Oh(m\log\log_w \sigma)$ within $\Oh(r\log n)$ bits, or
$\ftSearch = \Oh(m)$ within $\Oh(r\log(n/r)\log n)$ bits, where $r$ is the
number of equal-letter runs in the Burrows-Wheeler Transform of $\vTColl$.
This also upper-bounds the number of areas $\SA[i..i+\ell]$
into which $\SA$ can be divided such that each area appears elsewhere shifted
by 1 \cite{MN05}. 

The second component is the grammar $\vGrm$ that generates $\DA[1..\vsTColl]$, 
which must be binary and balanced. Such grammars can be built so as to ensure
that their total size is $\Oh(r\log(n/r)\log n)$ bits \cite{GNP18arxiv}, which
is of the same order of the first component.

The third component are the lists $D_v$ of the distinct documents that appear
in the expansion of each nonterminal $v$ of $\vGrm$. These lists are stored in
ascending order to merge them easily. To reduce their size, the set of
sequences $D_1, \ldots, D_g$ are grammar-compressed as a whole in a new grammar
$\vGrm'$, ensuring that no nonterminal of $\vGrm'$ expands beyond a list
$D_v$. Each list $D_v$ can then be obtained in optimal time, $\Oh(|D_v|)$, from
a nonterminal of $\vGrm'$.

\subsection{Document listing}

Given a pattern $P[1..m]$, we use the $\CSA$ to find the range 
$[\vSPosPat.. \vEPosPat]$ where the occurrences of $P$ lie in the suffix array,
in time $\Oh(\ftSearch)$. We then find the maximal nodes of the parse tree of 
$\DA$ that cover $\DA[\vSPosPat.. \vEPosPat]$.
Finally, we decompress the lists of the nonterminals corresponding to those maximal nodes, and compute their union.

Since $\vGrm$ is binary and balanced, there are $\Oh(\lg n)$ maximal 
nonterminals that cover $\DA[sp..ep]$ in the 
parse tree. By storing the length to which each nonterminal of $\vGrm$ expands,
we can easily find those $\Oh(\log n)$ maximal nonterminals in time
$\Oh(\log n)$, by (virtually) descending in the parse tree from the initial 
symbol of $\vGrm$ towards the area $\DA[sp..ep]$.

To merge the $\Oh(\log n)$ lists of documents in ascending order, we use an 
atomic heap \cite{FW94} (see practical considerations in the next section).
This data structure performs \texttt{insert} and \texttt{extractmin} 
operations in constant amortized time, when storing $\Oh(\lg^2 \vsTColl)$ 
elements. We then insert the heads of the $\Oh(\lg \vsTColl)$ lists in the 
atomic heap, extract the minimum, and insert the next element of its list.
If we extract the same document many times, we report only one copy.
We then expand and merge the lists $D_{v_1}, \dots, D_{v_k}$ in time 
$\Oh(|D_{v_1}| + \dots + |D_{v_k}|)$.

Since each distinct document we report may appear in the $\Oh(\lg \vsTColl)$ 
lists, our document listing solution takes time 
$\Oh(\ftSearch + \ndoc \cdot \lg \vsTColl)$. By using the RLCSA that occupies
$\Oh(r\log(n/r)\log n)$ bits, the total time is 
$\Oh(\vsPat + \ndoc \cdot \lg \vsTColl)$.

\subsection{Plugging-in other indexes}

Our grammar-compressed $\DA$, without the lists $D_v$, can be used to replace
the CSA component that requires $\Theta((n\log n)/\ftLookup)$ bits to compute
a cell in time $\Theta(\ftLookup)$. These indexes actually access cells
$\SA[i]$ in order to obtain $\DA[i]$. Our grammar-compressed $\DA$ offers
$\Oh(\log n)$ access time within $O(r\log(n/r)\log n)$ bits of space.

Therefore, we can implement Sadakane's solution \cite{Sad07}, as well as ILCP
and PDL \cite{GHKKNPS17} all answering in time $\Oh(m+\ndoc\cdot\log n)$, and
replacing the $\Oh((n\log n)/\ftLookup)$ part of their $|\CSA|$ space by
$O(r\log(n/r)\log n)$ bits (which also accounts for the RLCSA variant that finds
$[sp..ep]$ in time $\Oh(m)$. We can also implement the brute-force solution
in time $O(m+\occ+\log n)$ and $O(r\log(n/r)\log n)$ bits by extracting 
the whole $\DA[sp..ep]$.

\section{Practical Considerations}

\subsection{Compressed suffix array}

We use a practical RLCSA \cite[called RLFM+ in there]{MNSV09} that uses 
$(r\log\sigma + 2r\log(n/r))(1+o(1))$ bits of space and offers search time 
$\ftSearch$ in $\Oh(m\log r) \subseteq \Oh(m\log n)$. Since we do not need
to compute cells of $\SA$ with this structure, we do not need to spend the
$\Oh((n\log n)/\ftLookup)$ bits, and as a result the contribution of the RLCSA 
to the total space is negligible.

\subsection{Grammar compressor}

We choose Re-Pair \cite{LM00} to obtain both $\vGrm$ and $\vGrm'$, since it
performs very well in practice. Re-Pair repeatedly replaces the most frequent 
pair of adjacent symbols with a new nonterminal, until every pair is unique. 
Upon ties in frequency, we give priority to the pairs whose symbols have been
generated earlier, which in practice yielded rather balanced grammars in all 
the cases we have tried.

Re-Pair yields a binary grammar, but the top-level is a sequence of terminals
and nonterminals. 
We then complete the grammar by artificially adding a parse tree
on top of the final sequence left by Re-Pair. To minimize the height of the
resulting grammar, we merge first the pairs of nonterminals with shorter
parse trees.

We store the $\vnrules$ grammar rules as an array $G$ taking $2\vnrules
\lg(\vnrules + \vnDocs)$ bits of space, so that if $A_i$ is the $i$th
nonterminal of the grammar, it holds that $A_i \rightarrow A_{G[2 i]} A_{G[2 i + 1]}$.

When building $\vGrm'$, we concatenate all the lists $D_v$ and separate them
with unique numbers larger than $d$, to ensure that Re-Pair will not produce
nonterminals that cross from one list to another. After running Re-Pair, we 
remove the separators
but do not complete the grammars, as all we need is to decompress any $D_v$ in 
optimal time. We represent all the reduced sets $D'_v$ as a sequence $D'$, 
marking the beginning of each set in a bitvector $B$. The beginning of $D_v'$
is found with operation $select(B,v)$, which finds the $v$th 1 in $B$. This
operation can be implemented in constant time using $o(|B|)$ further bits 
\cite{Cla96}.

\subsection{Sampling}

The largest component of our index is the set of compressed lists $D_v'$. To
reduce this space, we will store those lists only for some sampled nonterminals
$v$ of $\vGrm$. The list of a nonsampled nonterminal $v$ is then obtained by
merging those of the highest sampled descendants of $v$ in the parse tree, 
which yields a space/time tradeoff.

We use a strategy similar to PDL \cite{GHKKNPS17}, based on parameters $b$ and
$\beta$. We define a {\em sampled tree} by sampling some nodes from the parse
tree. First, no leaf $v$ of the sampled tree can have an expansion larger than 
$b$, so that we spend time $\Oh(b\log b)$ to obtain its sorted list directly 
from $\mathcal{G}$. To this aim, we sample
all the nonterminals $v$ of $\vGrm$ with parent $w$ such that $|D_v| \le b <
|D_w|$. Those are the leaves of the sampled tree, which form a partition of 
$\DA$.

Second, for any nonsampled node $v$ with $|D_v| > b$, we must be able to build 
$D_v$ by merging other precomputed lists of total length $\le \beta |D_v|$. 
This implies that generating $D_v$ costs $\Oh(\vStorFac\lg n)$ times 
more than having $D_v'$ stored and just decompressing it.

We first assume that the sampled tree contains all the ancestors of the 
sampled leaves and then proceed bottom-up in the sampled tree, removing some
nodes from it. Any node $v$ with parent $w$ and children $u_1, ..., u_k$
is removed if $\sum_{i=1}^{k} |D_{u_i}| \le \vStorFac \cdot |D_v|$. In this
case, the nodes $u_i$ become children of $w$ in the sampled tree.

At query time, if a node $v$ of interest is not sampled, we collect all the
lists of its highest sampled descendants. Therefore, on a parse tree of height
$h$ we may end up merging many more than the original $\Oh(h)$ lists 
$D_1,\ldots,D_k$, but 
have the guarantee that the merged lists add up to size at most $\beta \cdot 
(|D_1|+ \cdots + |D_k|)$. To merge the lists we use a classical
binary heap instead of an atomic heap, so the cost per merged element is 
$\Oh(\log n)$. 

We may then spend $k\cdot b\log b = \Oh(hb \log b)$ time in extracting and
sorting the lists $D_v$ of size below $b$. The other lists $D_v$ may lead to 
merging $\beta |D_v|$ elements. The total cost over the $k=\Oh(h)$ lists is 
then $\Oh(hb\log b + \beta (|D_1|+ \cdots + |D_k|) \log n) \subseteq 
 \Oh(hb\log b + \ndoc \cdot \beta h \log n)$. In terms of complexity, if we 
choose for example $b = \Oh(\log n/\lg\lg n)$,
$\beta=\Oh(1)$, and the grammar is balanced, $h=\Oh(\log n)$, then the total 
cost of merging is $\Oh(\ndoc \cdot \log^2 n)$.

\section{Experiments and Results}

We evaluate different variants of our indexes and compare them with the state 
of the art. We use the experimental framework proposed by Gagie et 
al.~\cite{GHKKNPS17}.

\subsection{Document collections}
To test various kinds of repetitiveness scenarios, we performed several experiments with real and synthetic datasets.
We used the same document collections tested by Gagie et al.~\cite{GHKKNPS17},
available at \url{https://jltsiren.kapsi.fi/rlcsa}.
Table \ref{tab:collections} summarizes some statistics on the collections and the patterns used in the queries.
\begin{table*}
	\centering
\small
	\begin{tabu}{l r r r r r r r r}
		\toprule
		\rowfont{\em} Collection	& Size	& RLCSA			& Docs	& Doc size	& Patterns	& Occs						& Doc occs				& Occs/doc \\
		& ($\vsTColl$)	& (bps)	& ($D$)	& ($\vsTColl/D$)	&			& ($\occ$)	& ($\ndoc$)	& ($\frac{\occ}{\ndoc}$) \\
		
		\midrule
		\multirow{3}{*}{\texttt{Page}}		& 110	& 0.18		& \num{60}		& \num{1919382}	& \num{7658}	& \num{781}		& \num{3}		& 242.75 \\
		& 641	& 0.11		& \num{190}		& \num{3534921}	& \num{14286}	& \num{2601}	& \num{6}		& 444.79 \\
		& 1037	& 0.13		& \num{280}		& \num{3883145}	& \num{20536}	& \num{2889}	& \num{7}		& 429.04 \\
		
		\cline{2-9}
		\multirow{3}{*}{\texttt{Revision}}	& 110	& 0.18		& \num{8834}	& \num{13005}	& \num{7658}	& \num{776}		& \num{371}		& 2.09 \\
		& 640	& 0.11		& \num{31208}	& \num{21490}	& \num{14284}	& \num{2592}	& \num{1065}	& 2.43 \\
		& 1035	& 0.13		& \num{65565}	& \num{16552}	& \num{20536}	& \num{2876}	& \num{1188}	& 2.42 \\
		
		\cline{2-9}
%		\multirow{2}{*}{\texttt{Influenza}}	& 137	& 5.52		& \num{100000}	& \num{1436}	& \num{1000}	& \num{24975}	& \num{18547}	& 1.35 \\
%		& 321	& 10.53		& \num{227356}	& \num{1480}	& \num{1000}	& \num{59997}	& \num{44012}	& 1.36 \\
		\multirow{2}{*}{\texttt{Influenza}}	& 137	& 0.32		& \num{100000}	& \num{1436}	& \num{269}	& \num{532739}	& \num{88525}	& 6.02 \\
		& 321	& 0.26		& \num{227356}	& \num{1480}	& \num{269}	& \num{1248428}	& \num{202437}	& 6.17 \\
		
%		\midrule
%		\multirow{2}{*}{\texttt{Enwiki}}	& 113	& 49.44		& \num{7000}	& \num{16932}	& \num{18935}	& \num{1904}	& \num{505}		& 3.77 \\
%		& 639	& 309.31	& \num{44000}	& \num{15236}	& \num{19628}	& \num{10316}	& \num{2856}	& 3.61 \\
%		& 1034	& 482.16	& \num{90000}	& \num{12050}	& \num{19805}	& \num{17092}	& \num{4976}	& 3.44 \\
%
%		\multirow{1}{*}{\texttt{Swissprot}}	& 54	& 25.19		& \num{143244}	& \num{398}		& \num{10000}	& \num{160}		& \num{121}		& 1.33 \\
		
		\midrule
%		\multirow{1}{*}{\texttt{DNA}}	    & 95	&	& \num{100000}					&	& \num{889}\text{--}\num{1000}		&	&	& \\
%		\multirow{1}{*}{\texttt{Concat}}	& 95	&	& \num{10}\text{--}\num{100}	&	& \num{7538}\text{--}\num{13847}	&	&	& \\
%		\multirow{1}{*}{\texttt{Version}}	& 95	&	& \num{10000}					&	& \num{7537}\text{--}\num{13846}	&	&	& \\
		\bottomrule
	\end{tabu}
	
	\caption{
		Statistics for document collections (small, medium, and large variants):
		\emph{Collection} name; \emph{Size} in megabytes; \emph{RLCSA} bits per symbol (bps); \emph{Docs}, number of documents; \emph{Doc size}, average document length; number of \emph{Patterns}; \emph{Occs}, average number of occurrences; \emph{Doc occs}, average number of document occurrences; \emph{Occs/doc}, average ratio of occurrences to document occurrences.
	}
	
	\label{tab:collections}
\end{table*}

\subparagraph{Real collections. \rm
\texttt{Page} and \texttt{Revision} are collections formed by all the revisions of some selected pages from the Wikipedia in Finnish language.
In \texttt{Page}, there is a document for each selected article, that also includes all of its revisions.
In the case of \texttt{Revision}, each page revision becomes a separate document.
\texttt{Influenza} is another repetitive collection composed of sequences of
the H.\ influenzae virus genomes. }
%\textcolor{red}{Add other collection (code repository) -- no es tan importante}

%\paragraph{Non-repetitive collections.}
%\texttt{Enwiki} is a non-repetitive collection of pages from a snapshot of the English-language Wikipedia.
%Unlike \texttt{Page} and \texttt{Revision}, \texttt{Enwiki} does not include the different versions of each article.
%The other non-repetitive dataset tested is \texttt{Swissprot} which is formed by protein sequences.

\subparagraph{Synthetic collections. \rm
We also used two types of synthetic collections to explore the effect of collection repetitiveness on document listing performance in more detail.
\texttt{Concat} and \texttt{Version} are similar to \texttt{Page} and \texttt{Revision}, respectively.
We use 10 and 100 base documents of length $\num{1000}$ each, extracted at random from the English file of Pizza\&Chili ({\tt http://pizzachili.dcc.uchile.cl}).
Besides, we include variants of each base document, generated using different
mutation probabilities ($\num{0.001}$, $\num{0.003}$, $\num{0.01}$,
and $\num{0.03}$). A mutation is a replacement by a different random symbol.
In collection \texttt{Version}, each variant becomes a separate document.
In \texttt{Concat}, all variants of the same base document are concatenated
into a single document. }

\subparagraph{Queries. \rm
The query patterns for \texttt{Page} and \texttt{Revision} datasets are Finnish words of length $\geq 5$ that occur in the collections.
For \texttt{Influenza}, the queries are substrings of length $4$ extracted from the dataset.
In the case of \texttt{Concat} and \texttt{Version}, the patterns are terms selected from an MSN query log.
See Gagie et al.~\cite{GHKKNPS17} for a more detailed description.}

\subsection{Compared indexes}

\subparagraph*{Grammar-Compressed Document Array (\emph{GCDA}). \rm 
This is our main proposal. We use the balanced variant of the Re-Pair compressor implemented by Navarro\footnote{\url{https://www.dcc.uchile.cl/gnavarro/software/repair.tgz}}.  To sample the parse tree, we test several parameter configurations for the block size $\vsBlock$ and factor $\vStorFac$.}

\subparagraph*{Brute force (\emph{Brute}). \rm
This family of algorithms is the most basic and simple solution to the document listing problem.
They use a $\CSA$ to retrieve all the document identifiers in $\DA[\vSPosPat.. \vEPosPat]$, sort them, and report each of them once.
\idxName{Brute-L} uses the $\CSA$ to extract the values $\DA[i]$.
\idxName{Brute-D}, instead, uses an explicit document array $\DA$. Finally,
\idxName{Brute-C} is our variant using the grammar-compressed $\DA$.
From the grammar tree of height $h$ and storing the length of the expansion of 
each nonterminal, we extract the range $\DA[sp..ep]$ in time $\Oh(h+ep-sp)$.}

\subparagraph*{Sadakane (\emph{Sada}). \rm
\idxName{Sada-L} is the original index proposed by Sadakane \cite{Sad07}.
\idxName{Sada-D} speeds up the query time by explicitly storing 
$\DA$. \idxName{Sada-C} stores $\DA$ in grammar-compressed form, where each 
individual cell $\DA[i]$ is extracted in time $\Oh(h)$.}

\subparagraph*{Interleaved Longest Common Prefix (\emph{ILCP}). \rm
\idxName{ILCP-L} is an implementation of the ILCP index proposed by Gagie et
al.~\cite{GHKKNPS17} using a run-length encoded \idxName{ILCP} array.
\idxName{ILCP-D} is a variant that uses the document array instead of the 
$\CSA$ functionality. \idxName{ILCP-C} uses, instead, our grammar-compressed
$\DA$, which accesses any cell in time $\Oh(h)$.}

\subparagraph*{Precomputed Document Lists (\emph{PDL}). \rm
\idxName{PDL-BC} and \idxName{PDL-RP} are implementations of the PDL algorithm 
proposed by Gagie et al.~\cite{GHKKNPS17}.
The first one uses a Web graph compressor \cite{HN13} on the set of lists, 
whereas \idxName{PDL-RP} uses Re-Pair compression. Both variants use block size 
$\vsBlock = 256$ and factor $\vStorFac = 16$, as recommended by their
authors.}

\subparagraph*{Grammar-based (\emph{Grammar}). \rm
This is an implementation of the index by Claude and Munro 
\cite{CM13}. It uses Re-Pair on the collection 
$\vTColl$ and on the set of lists. This index is the only tested solution that 
does not use a $\CSA$.}

We implemented \idxName{GCDA} on C++, using several succinct data structures from the \texttt{SDSL} library\footnote{\url{https://github.com/simongog/sdsl-lite}}.
We used existing C++ implementations of the indexes \idxName{Brute}, 
\idxName{Sada}, \idxName{ILCP} and \idxName{PDL}, which were tested by 
Gagie et al.~\cite{GHKKNPS17}%
\footnote{\url{https://jltsiren.kapsi.fi/software/doclist.tgz}}, and 
modified the versions \idxName{-C} by using $\DA$ in grammar-compressed instead
of in plain form.

All tested indexes except \idxName{Grammar} use a suffix array to compute the 
interval $[sp..ep]$ corresponding to pattern $P$.
We used a RLCSA implementation\footnote{\url{https://jltsiren.kapsi.fi/rlcsa}} 
that is optimized for highly repetitive text collections.
To compute entries $\SA[i]$, the RLCSA uses a suffix array sampling, which
requires significant space as explained.
Our index does not use this operation, but it is required for the indexes \idxName{Brute-L}, \idxName{Sada-L}, \idxName{ILCP-L}, and both variants of \idxName{PDL}.
We use 32 as the value for this sample rate, as it gave good results in previous
tests \cite{GHKKNPS17}. The exception is \idxName{Brute-L}, which uses a
RLCSA optimized to extract whole ranges $\SA[sp..ep]$ \cite{GNP18}.%
\footnote{\tt https://github.com/nicolaprezza/r-index}
The column {\em RLCSA} of 
Table~\ref{tab:collections} gives the space used by the RLCSA 
without suffix array samples.

Our machine has two Intel(R) Xeon(R) CPU E5-2407 processors running at $\SI{2.40}{\GHz}$ and $\SI{250}{\gibi\byte}$ of RAM.
The operating system was Debian with Linux kernel \texttt{4.9.0-8-amd64}.
All indexes were compiled using \texttt{g++} version \texttt{6.3.0} with flags \texttt{-O3 -DNDEBUG}.

%To better appreciate the performance differences between the indexes, we omit the search time for the interval $[\vSPosPat.. \vEPosPat]$.
%This common step takes approximately $\num{3}\text{--}\num{5}$ microseconds per pattern.
%Only index \idxName{Grammar} does not need this operation, but since it requires at least $\num{570}$ microseconds, it is not noticeably harmed.
%\textcolor{red}{No seria mejor sumarles los 3--5 microsec a los otros, en vez
%de dar tantas explicaciones?}

\subsection{Tuning our main index}

%Table \todo{Ref} shows the detailed results obtain in our experiments.
%We show in terms of space and the query-time required by each index.

\begin{figure}[t]
	\centering
\hspace*{-8mm}
	\begin{subfigure}[b]{0.36\linewidth}
		\includegraphics[width=\textwidth]{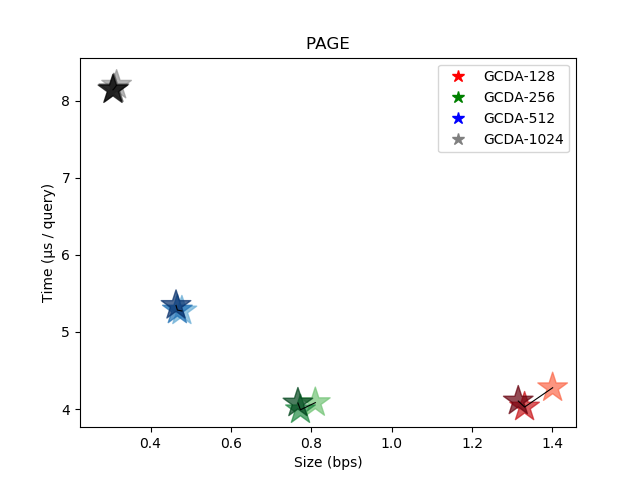}
%		\caption{10 base documents}
%		\label{fig:gull}
	\end{subfigure}
%	~ %add desired spacing between images, e. g. ~, \quad, \qquad, \hfill etc.
	%(or a blank line to force the subfigure onto a new line)
\hspace*{-5mm}
	\begin{subfigure}[b]{0.36\linewidth}
		\includegraphics[width=\textwidth]{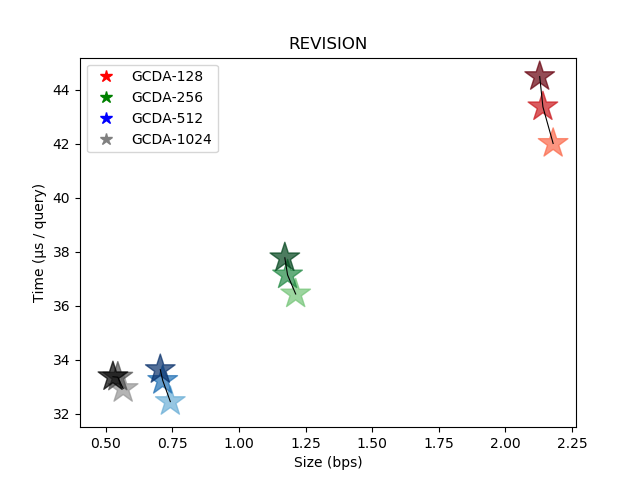}
%		\caption{100 base document}
%		\label{fig:tiger}
	\end{subfigure}
%	~ %add desired spacing between images, e. g. ~, \quad, \qquad, \hfill etc.
	%(or a blank line to force the subfigure onto a new line)
\hspace*{-3mm}
	\begin{subfigure}[b]{0.36\linewidth}
	\includegraphics[width=\textwidth]{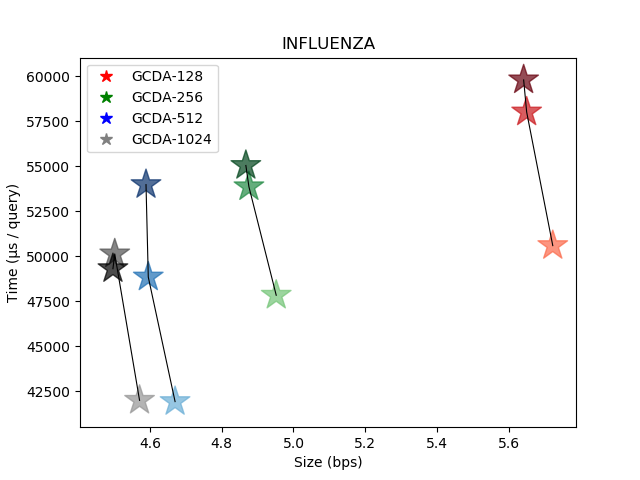}
	%		\caption{100 base document}
	%		\label{fig:tiger}
	\end{subfigure}
\hspace*{-8mm}

	\caption{\idxName{GCDA} on small real collections with different configurations. The $x$ axis shows the total size of the index in bits per symbol. The $y$ axis shows the average time per query in microseconds. Beware that the 
plots do not start at zero.}
	\label{fig:coll_real_tradeoff}
\end{figure}

Figure \ref{fig:coll_real_tradeoff} shows the tradeoff between time and space 
of \idxName{GCDA} on small real collections.
We tested \idxName{GCDA} with 4 different sizes of block $\vsBlock$: $\num{128}$, $\num{256}$, $\num{512}$, and $\num{1024}$.
For each block size, we used 3 different factors $\vStorFac$ ($\num{4}$, $\num{8}$, and $\num{16}$), which are represented with increasing color darkness in the plots.
The configuration $\vsBlock = 512$ and $\vStorFac = 4$ shows to be a good general-purpose choice of parameter values, and we stick to it from now on.

The lower-right plot of Figure~\ref{fig:repetitive_collections} shows the
space required
by the main components of our index. As the number of documents in the 
collection grows and their size decreases, the weight of the grammar-compressed
$\DA$, and even more, of the grammar-compressed lists of documents, becomes
dominant. Note also that \texttt{Influenza} is the least repetitive collection.

\subsection{Comparison on real collections}

\begin{figure*}[t!]
%	\centering
	\hspace*{-1mm}
	\begin{subfigure}[b]{0.36\linewidth}
		\includegraphics[width=\textwidth]{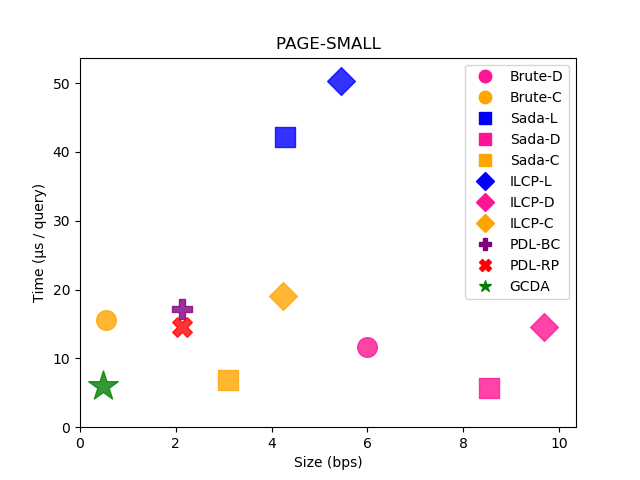}
%		\caption{small}
%		\label{fig:gull}
	\end{subfigure}
	~ %add desired spacing between images, e. g. ~, \quad, \qquad, \hfill etc.
	%(or a blank line to force the subfigure onto a new line)
	\hspace*{-9mm}
	\begin{subfigure}[b]{0.36\linewidth}
		\includegraphics[width=\textwidth]{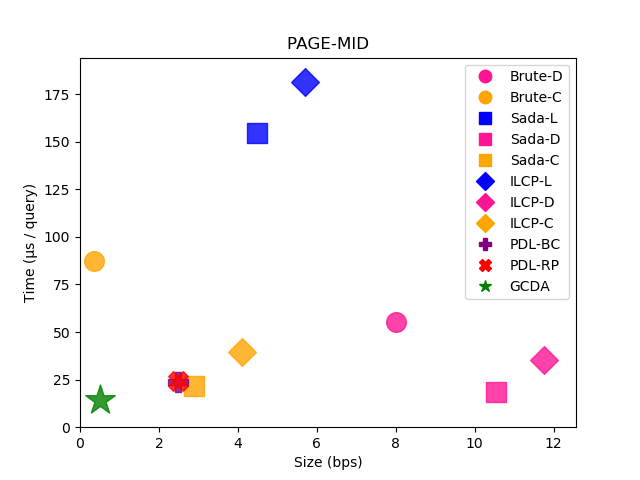}
%		\caption{medium}
%		\label{fig:tiger}
	\end{subfigure}
	~ %add desired spacing between images, e. g. ~, \quad, \qquad, \hfill etc.
	%(or a blank line to force the subfigure onto a new line)
	\hspace*{-9mm}
	\begin{subfigure}[b]{0.36\linewidth}
		\includegraphics[width=\textwidth]{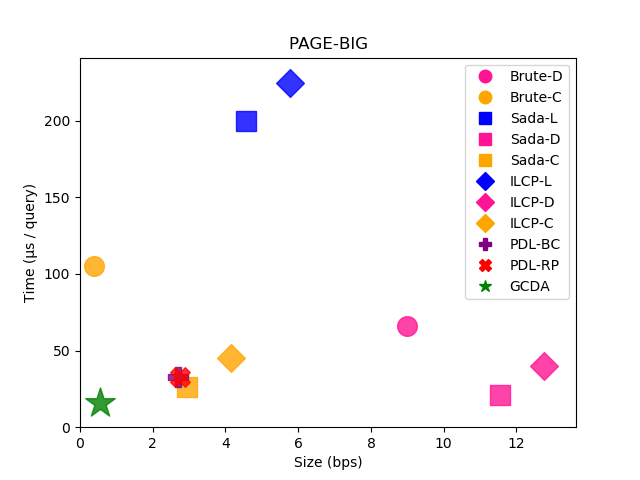}
%		\caption{large}
%		\label{fig:mouse}
	\end{subfigure}
	\hspace*{-10mm}

	\hspace*{-1mm}
	\begin{subfigure}[b]{0.36\linewidth}
		\includegraphics[width=\textwidth]{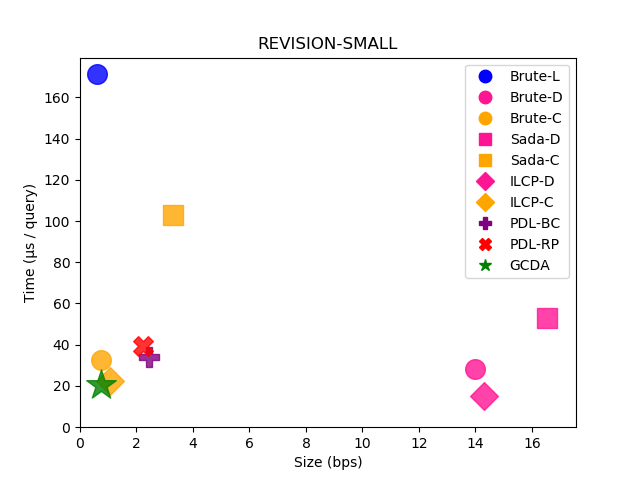}
%		\caption{small}
%		\label{fig:gull}
	\end{subfigure}
	~ %add desired spacing between images, e. g. ~, \quad, \qquad, \hfill etc.
	%(or a blank line to force the subfigure onto a new line)
	\hspace*{-9mm}
	\begin{subfigure}[b]{0.36\linewidth}
		\includegraphics[width=\textwidth]{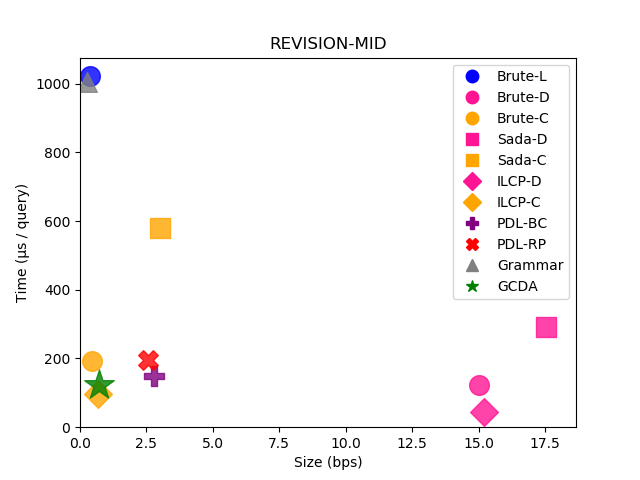}
%		\caption{medium}
%		\label{fig:tiger}
	\end{subfigure}
	~ %add desired spacing between images, e. g. ~, \quad, \qquad, \hfill etc.
	%(or a blank line to force the subfigure onto a new line)
	\hspace*{-9mm}
	\begin{subfigure}[b]{0.36\linewidth}
		\includegraphics[width=\textwidth]{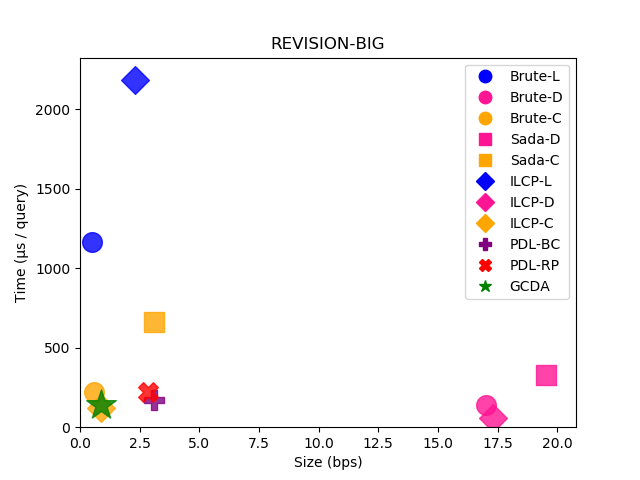}
%		\caption{large}
%		\label{fig:mouse}
	\end{subfigure}
	\hspace*{-10mm}

	\hspace*{-1mm}
	\begin{subfigure}[b]{0.36\linewidth}
		\includegraphics[width=\textwidth]{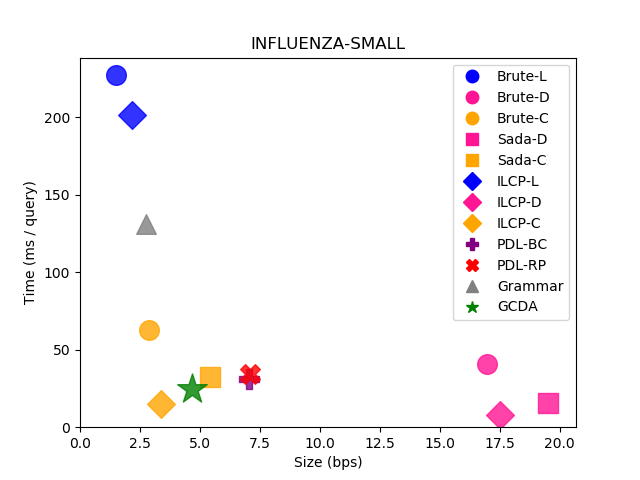}
%		\caption{small}
%		\label{fig:gull}
	\end{subfigure}
	~ %add desired spacing between images, e. g. ~, \quad, \qquad, \hfill etc.
	%(or a blank line to force the subfigure onto a new line)
	\hspace*{-9mm}
	\begin{subfigure}[b]{0.36\linewidth}
		\includegraphics[width=\textwidth]{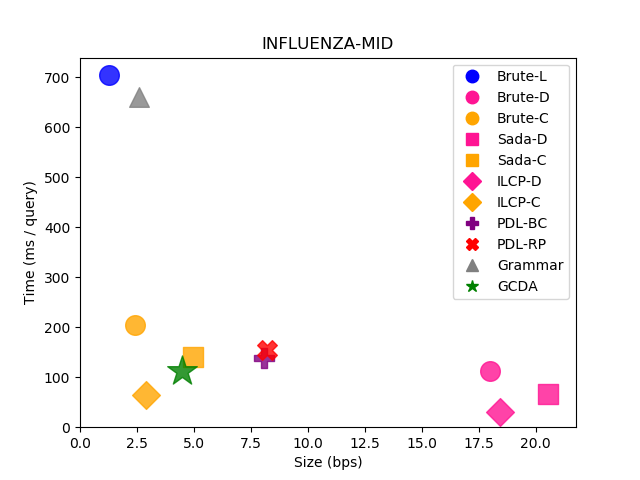}
%		\caption{medium}
%		\label{fig:tiger}
	\end{subfigure}
	~ %add desired spacing between images, e. g. ~, \quad, \qquad, \hfill etc.
	%(or a blank line to force the subfigure onto a new line)
	\hspace*{-9mm}
	\begin{subfigure}[b]{0.36\linewidth}
%		\vspace*{-9mm}
		\includegraphics[width=\textwidth]{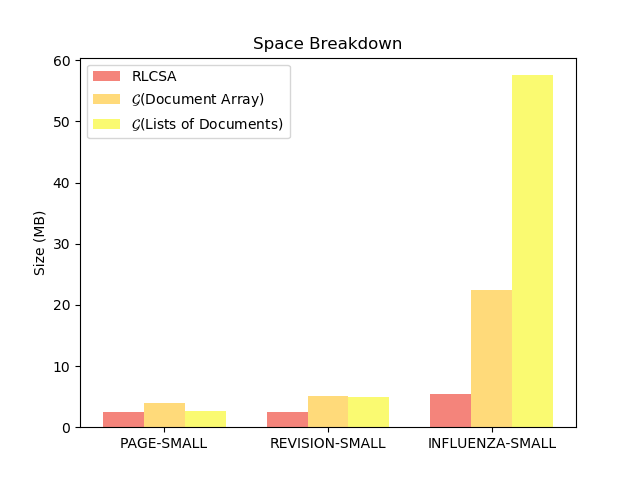}
%		\caption{large}
%		\label{fig:mouse}
	\end{subfigure}
    \hspace*{-10mm}

	\caption{Document listing indexes on real repetitive collections. The $x$ axis shows the total size of the index in bits per symbol. The $y$ axis shows the average time per query. Combinations with excessively high
time are omitted in some plots. The lower-right plot shows the size of the
main components of \idxName{GCDA} on the small collections;
the $y$ axis shows the size in megabytes.
	}
	\label{fig:repetitive_collections}
\end{figure*}

Figure~\ref{fig:repetitive_collections} shows the tradeoff between time and 
space for all tested indexes on the real collections. Our main index, 
\idxName{GCDA}, and the \idxName{-C} variants of the other indexes we adapted,
are clearly dominant in a large portion of the space/time map. Most of the
previous indexes are way slower, way larger, or both, than ours. The best
previous tradeoffs, \idxName{PDL-BC} and \idxName{PDL-RP} \cite{GHKKNPS17}, are
much closer, but still they are almost always slower and larger than
\idxName{GCDA}.

For all versions of \texttt{Page}, where there are few large documents and our 
grammars compress very well, \idxName{GCDA} requires only 
$\num{0.48}\text{--}\num{0.56}$ bits per symbol (bps) and answers queries in
less than 16 microseconds. The index using the
least space is \idxName{Grammar}, which requires $\num{0.21}\text{--}\num{0.35}$
bps. \idxName{Grammar} is way out of the plot, however, because it requires 
1.2--3.4 milliseconds to solve the queries, that is, 205--235 times slower 
than \idxName{GCDA}.\footnote{As in previous work \cite{GHKKNPS17}, 
\idxName{Grammar} was not built on the largest dataset of \texttt{Page}.} 
The next smallest index is our variant \idxName{Brute-C}, which uses
0.35--0.55 bps and is generally smaller than \idxName{GCDA}, but slower by
a factor of 2.6--6.7. \idxName{Brute-L}, occupying 0.38--0.60 bps, is also
smaller in some cases, but much slower (180--1080 microseconds, out of the
plot). \idxName{GCDA} sharply outperforms all the other indexes in
space, and also in time (only \idxName{Sada-D} is 6\% faster
in the small collection, yet using 18 times more space). The closest
competitors, \idxName{PDL-BC} and \idxName{PDL-RP}, are 4.4--5.0
times larger and $\num{2.8}\text{--}\num{5.0}$ times slower than 
\idxName{GCDA}.

In the case of \texttt{Revision}, where there are more and smaller documents,
\idxName{GCDA} uses 0.73--0.88 bps and answers queries in less than 150
microseconds. Again \idxName{Grammar} uses the least
space, 0.26--0.42 bps, but once again at the price of being 
$\num{8}\text{--}\num{30}$ times slower than \idxName{GCDA}. 
The case of \idxName{Brute-L} is
analogous: 0.38--0.60 bps but over 8 times slower than \idxName{GCDA}.
Instead, our variant \idxName{Brute-C} is a relevant competitor, using
0.45--0.76 bps and being less than 60\% slower than \idxName{GCDA}. The other
relevant index is our variant \idxName{ILCP-C}, using almost the same space
and time of \idxName{GCDA}. The group \idxName{GCDA/Brute-C/ILCP-C} forms a
clear sweetpoint in this collection. The closest competitors, again 
\idxName{PDL-BC} and \idxName{PDL-RP}, are 3.1--3.8 times larger
and 1.2--1.9 times slower than \idxName{GCDA}.

\texttt{Influenza}, with many small documents, is the worst case for the
indexes. \idxName{GCDA} uses 4.46--4.67 bps and answers queries within 115
milliseconds. Many indexes are smaller than \idxName{GCDA}, but only our
variants form a relevant space/time tradeoff: \idxName{ILCP-C} uses 2.88--3.37
bps, \idxName{Brute-C} uses 2.42--2.86 bps, and \idxName{Sada-C} uses
4.96--5.40 bps. All the \idxName{-C} variants obtain competitive times, and
\idxName{ILCP-C} even dominates \idxName{GCDA} (it answers queries within 65
milliseconds, taking less than 60\% of the time of \idxName{GCDA}). The other 
indexes outperforming \idxName{GCDA} in time are \idxName{-D} variants, which 
are at least 3.7 times larger than \idxName{GCDA} and 5.2 times larger than
\idxName{ILCP-C}.

%\todo[disable, inline]{Edit}
%In general, collection size is more important in top-k document retrieval. Increasing the number of documents generally increases the df/k ratio, and thus makes brute-force solutions based on docu- ment listing less appealing.
%In document listing, the size of the documents is more important than collection size, as a large occ/df ratio makes brute-force solutions based on pattern matching less appealing.

%\paragraph{Non-repetitive Collections.}
%Figure \ref{fig:non-repetitive_collections} contains the results for non-repetitive collections.
%\input{./tex/img_nonrep_coll_res}

%On \texttt{Enwiki} and \texttt{Swissprot}, our index achieves a very good performance in the query-time, but the cost in space is very high.
%For small version of both collections, \idxName{GCDA} is the second fastest index, only outperformed by \idxName{Brute-D}.

%\idxName{PDL-BC} outperforms all indexes on this kind of collections, achieving a very good time/space balance.

\subsection{Comparison on synthetic collections}

\begin{figure}[t]
	\centering
	\begin{subfigure}[b]{0.49\linewidth}
		\includegraphics[width=\textwidth]{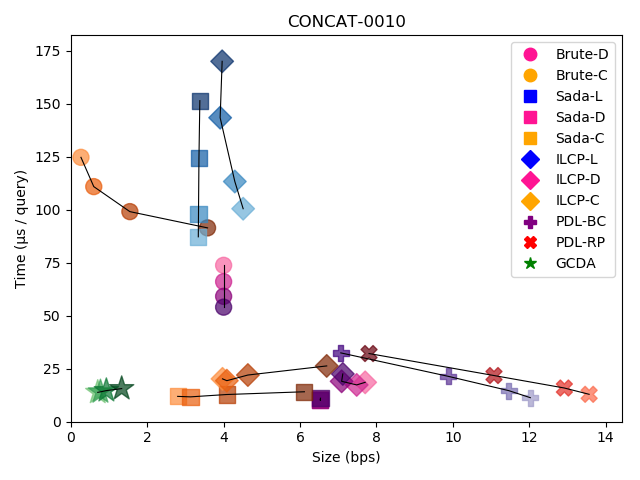}
%		\caption{10 base documents}
%		\label{fig:gull}
	\end{subfigure}
%	~ %add desired spacing between images, e. g. ~, \quad, \qquad, \hfill etc.
	%(or a blank line to force the subfigure onto a new line)
	\begin{subfigure}[b]{0.49\linewidth}
		\includegraphics[width=\textwidth]{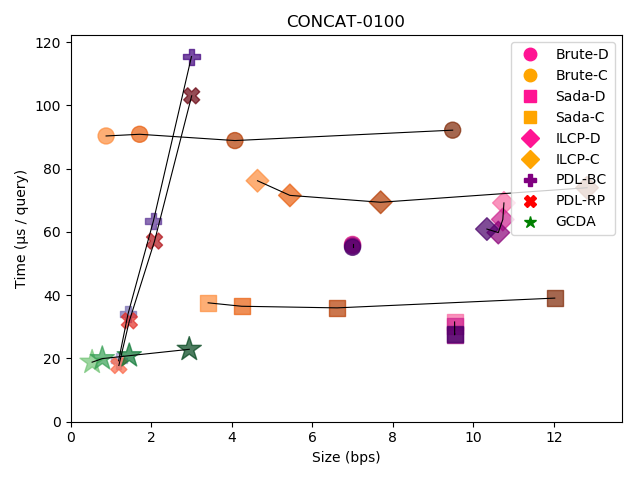}
%		\caption{100 base document}
%		\label{fig:tiger}
	\end{subfigure}

	\centering
	\begin{subfigure}[b]{0.49\linewidth}
		\includegraphics[width=\textwidth]{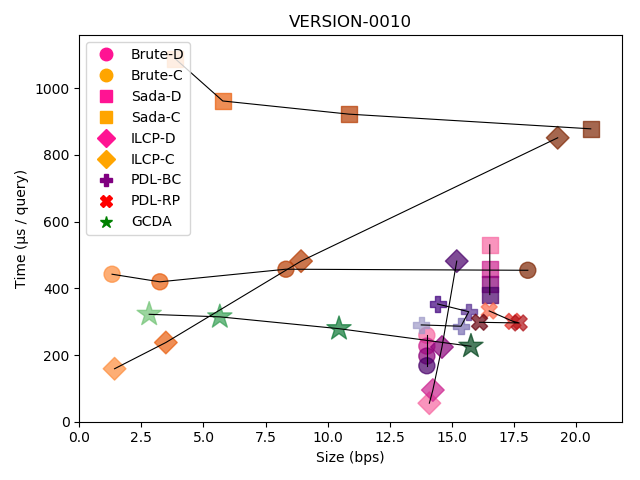}
%		\caption{10 base documents}
%		\label{fig:gull}
	\end{subfigure}
%	~ %add desired spacing between images, e. g. ~, \quad, \qquad, \hfill etc.
	%(or a blank line to force the subfigure onto a new line)
	\begin{subfigure}[b]{0.49\linewidth}
		\includegraphics[width=\textwidth]{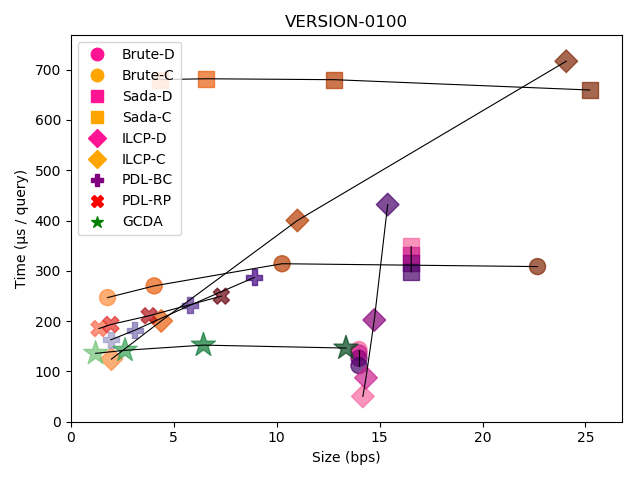}
%		\caption{100 base document}
%		\label{fig:tiger}
	\end{subfigure}

	\caption{
		Document listing on synthetic collections. The $x$ axis shows the total size of the index in bits per symbol. The $y$ axis shows the average time per query in microseconds. Combinations with excessively high time are
omitted in some plots. 
%		We omit the results for the slowest indexes.
	}
	\label{fig:coll_synthetic}
\end{figure}

Figure \ref{fig:coll_synthetic} compares the indexes on synthetic collections.
These allow us study how the indexes evolve as the
repetitiveness decreases, in a scenario of few large documents 
(\texttt{Concat}) and many smaller documents (\texttt{Version}). 
We combine in a single plot the results for different mutation rates of a given
collection and number of base documents. The plots show the increasing mutation
rates using variations of the same color, from lighter to darker. All the 
\idxName{-L} variants and \idxName{Grammar} are omitted because they were 
significantly slower.

On collection \texttt{Concat}, \idxName{GCDA} essentially outperforms all the other 
indexes. In the case of the version composed by $\num{10}$ base documents, 
our index obtains the best space/time tradeoff by a wide margin. Only
\idxName{Brute-C} is smaller than \idxName{GCDA}, but 8--9 times slower. On
the other hand, various indexes are slightly faster than \idxName{GCDA}, but 
much larger (from \idxName{Sada-D}, which is up to 30\% faster but 7 times 
larger, to \idxName{Sada-C}, which is 15\% faster but at least 4 times larger).
With the other variant of \texttt{Concat} ($\num{100}$ base documents), our 
index offers the best space and time for all mutation rates. Only
\idxName{PDL-RP} is 6\% faster in its best case, but 2.2 times larger.
Further, \idxName{GCDA} retains its space/time performance as repetitiveness
decreases, whereas the competing indexes worsen fast in one or both aspects.

On \texttt{Version}, composed by $\num{10000}$ documents of length
$\num{1000}$, \idxName{GCDA} is also a dominant solution, retaining its time
performance as repetitiveness decreases and outperforming all the \idxName{-D}
variants in space up to a mutation rate of 1\%. Other competing indexes are our 
variants \idxName{Brute-C}
and \idxName{ILCP-C} (the only one dominating \idxName{GCDA} in some cases), 
as well as \idxName{PDL-BC} and \idxName{PDL-RP} in the
case of 100 base documents. The strange behavior of the \idxName{PDL} indexes 
in both collections with 10 base documents is briefly discussed in the original
article \cite{GHKKNPS17}.

\section{Conclusions}

We have presented simple and efficient indexes for document listing on 
repetitive string collections. They find the $\ndoc$ documents where a pattern
of length $m$ appears in a collection of size $n$ in time 
$\Oh(\vsPat + \ndoc \cdot \lg \vsTColl)$. The indexes uses grammar-compression
of the document array, and perform better as the collection is more repetitive.

Our experimental results show that our main index, \idxName{GCDA}, outperforms 
the best previous solutions by a fair margin in time and/or space on various 
repetitive collections. From the previous indexes, only \idxName{PDL} 
\cite{GHKKNPS17} gets close, but it is almost always dominated by 
\idxName{GCDA} in both space and time. \idxName{GCDA}
performs well in space for mutation rates up to 1\%,
whereas its query time is mostly insensitive to the repetitiveness.
Other previous solutions (especially \idxName{ILCP} \cite{GHKKNPS17} and brute
force) that we adapted to run on our 
grammar-compressed document array also display unprecedented performance on 
repetitive texts, competing with \idxName{GCDA}.

For the final version of this paper, we plan to combine the \idxName{PDL}
indexes with a grammar-compressed document array as well, which we omitted for 
lack of time.  A line of future work is to further reduce the space of 
\idxName{GCDA} and our index variants that use the grammar-compressed document
array, by using a more clever encoding of the grammars that may nearly halve 
their space at a
modest increase in time \cite{GNF14}. Another line is to extend the index to
support {\em top-$k$ document retrieval}, that is, find the $k$ documents where
$P$ appears most often. For example, following previous ideas
\cite{GHKKNPS17}, we can store the list of documents where each nonterminal
appears in decreasing order of frequency, and use algorithms developed for
inverted indexes \cite{BYRN11} on the $\Oh(\log n)$ lists involved in a query.
The frequency of the candidates can be efficiently counted on repetitive
collections \cite{GHKKNPS17}.

% JOURNAL O FINAL: Tambien PDL puede usar el DA!

\newpage

\bibliography{paper}

\begin{thebibliography}{10}

\bibitem{BYRN11}
R.~Baeza-Yates and B.~Ribeiro-Neto.
\newblock {\em Modern Information Retrieval}.
\newblock Addison-Wesley, 2nd edition, 2011.

\bibitem{BN13}
D.~Belazzougui and G.~Navarro.
\newblock Alphabet-independent compressed text indexing.
\newblock {\em ACM Transactions on Algorithms}, 10(4):article 23, 2014.

\bibitem{CLLPPSS05}
M.~Charikar, E.~Lehman, D.~Liu, R.~Panigrahy, M.~Prabhakaran, A.~Sahai, and
  A.~Shelat.
\newblock The smallest grammar problem.
\newblock {\em IEEE Transactions on Information Theory}, 51(7):2554--2576,
  2005.

\bibitem{Cla96}
D.~R. Clark.
\newblock {\em Compact {PAT} Trees}.
\newblock PhD thesis, University of Waterloo, Canada, 1996.

\bibitem{CM13}
F.~Claude and J.~I. Munro.
\newblock Document listing on versioned documents.
\newblock In {\em Proc. 20th Symposium on String Processing and Information
  Retrieval (SPIRE)}, LNCS 8214, pages 72--83, 2013.

\bibitem{CN10}
F.~Claude and G.~Navarro.
\newblock Self-indexed grammar-based compression.
\newblock {\em Fundamenta Informaticae}, 111(3):313--337, 2010.

\bibitem{FW94}
M.~L. Fredman and D.~E. Willard.
\newblock {Trans-dichotomous algorithms for minimum spanning trees and shortest
  paths}.
\newblock {\em Journal of Computer and System Sciences}, 48(3):533--551, 1994.

\bibitem{GHKKNPS17}
T.~Gagie, A.~Hartikainen, K.~Karhu, J.~K{\"a}rkk{\"a}inen, G.~Navarro, S.~J.
  Puglisi, and J.~Sir{\'e}n.
\newblock Document retrieval on repetitive collections.
\newblock {\em Information Retrieval}, 20:253--291, 2017.

\bibitem{GNP18arxiv}
T.~Gagie, G.~Navarro, and N.~Prezza.
\newblock Fully-functional suffix trees and optimal text searching in
  {BWT}-runs bounded space.
\newblock {\em CoRR}, abs/1809.02792, 2018.

\bibitem{GNP18}
T.~Gagie, G.~Navarro, and N.~Prezza.
\newblock Optimal-time text indexing in {BWT}-runs bounded space.
\newblock In {\em Proc. 29th Annual ACM-SIAM Symposium on Discrete Algorithms
  (SODA)}, pages 1459--1477, 2018.

\bibitem{GNF14}
R.~Gonz{\'a}lez, G.~Navarro, and H.~Ferrada.
\newblock Locally compressed suffix arrays.
\newblock {\em ACM Journal of Experimental Algorithmics}, 19(1):article 1,
  2014.

\bibitem{HN13}
C.~Hern{\'a}ndez and G.~Navarro.
\newblock Compressed representations for web and social graphs.
\newblock {\em Knowledge and Information Systems}, 40(2):279--313, 2014.

\bibitem{Jez16}
A.~Jez.
\newblock A really simple approximation of smallest grammar.
\newblock {\em Theoretical Computer Science}, 616:141--150, 2016.

\bibitem{KN13}
S.~Kreft and G.~Navarro.
\newblock On compressing and indexing repetitive sequences.
\newblock {\em Theoretical Computer Science}, 483:115--133, 2013.

\bibitem{LM00}
J.~Larsson and A.~Moffat.
\newblock Off-line dictionary-based compression.
\newblock {\em Proceedings of the IEEE}, 88(11):1722--1732, 2000.

\bibitem{LS02}
E.~Lehman and A.~Shelat.
\newblock Approximation algorithms for grammar-based compression.
\newblock In {\em Proc. 13th Annual ACM-SIAM Symposium on Discrete Algorithms
  (SODA)}, pages 205--212, 2002.

\bibitem{MN05}
V.~M{\"a}kinen and G.~Navarro.
\newblock Succinct suffix arrays based on run-length encoding.
\newblock {\em Nordic Journal of Computing}, 12(1):40--66, 2005.

\bibitem{MNSV09}
V.~M{\"a}kinen, G.~Navarro, J.~Sir{\'e}n, and N.~V{\"a}lim{\"a}ki.
\newblock Storage and retrieval of highly repetitive sequence collections.
\newblock {\em Journal of Computational Biology}, 17(3):281--308, 2010.

\bibitem{MM93}
U.~Manber and G.~Myers.
\newblock Suffix arrays: a new method for on-line string searches.
\newblock {\em SIAM Journal on Computing}, 22(5):935--948, 1993.

\bibitem{Mut02}
S.~Muthukrishnan.
\newblock Efficient algorithms for document retrieval problems.
\newblock In {\em Proc. 13th Annual ACM-SIAM Symposium on Discrete Algorithms
  (SODA)}, pages 657--666, 2002.

\bibitem{Nav12}
G.~Navarro.
\newblock Indexing highly repetitive collections.
\newblock In {\em Proc. 23rd International Workshop on Combinatorial Algorithms
  (IWOCA)}, LNCS 7643, pages 274--279, 2012.

\bibitem{Nav14}
G.~Navarro.
\newblock Spaces, trees and colors: The algorithmic landscape of document
  retrieval on sequences.
\newblock {\em ACM Computing Surveys}, 46(4):article 52, 2014.

\bibitem{Nav17}
G.~Navarro.
\newblock Document listing on repetitive collections with guaranteed
  performance.
\newblock In {\em Proc. 28th Annual Symposium on Combinatorial Pattern Matching
  (CPM)}, LIPIcs 78, page article 4, 2017.

\bibitem{NM06}
G.~Navarro and V.~M{\"a}kinen.
\newblock Compressed full-text indexes.
\newblock {\em ACM Computing Surveys}, 39(1):article 2, 2007.

\bibitem{Ryt03}
W.~Rytter.
\newblock Application of {L}empel-{Z}iv factorization to the approximation of
  grammar-based compression.
\newblock {\em Theoretical Computer Science}, 302(1-3):211--222, 2003.

\bibitem{Sad07}
K.~Sadakane.
\newblock Succinct data structures for flexible text retrieval systems.
\newblock {\em Journal of Discrete Algorithms}, 5:12--22, 2007.

\bibitem{Plos15}
Z.~D. Sthephens, S.~Y. Lee, F.~Faghri, R.~H. Campbell, Z.~Chenxiang, M.~J.
  Efron, R.~Iyer, S.~Sinha, and G.~E. Robinson.
\newblock Big data: Astronomical or genomical?
\newblock {\em PLoS Biology}, 17(7):e1002195, 2015.

\bibitem{Wei73}
P.~Weiner.
\newblock Linear pattern matching algorithm.
\newblock In {\em Proc. 14th Annual IEEE Symposium on Switching and Automata
  Theory}, pages 1--11, 1973.

\end{thebibliography}

\end{document}